\documentclass[aps,pra,superscriptaddress,twocolumn]{revtex4}
\usepackage[T1]{fontenc}
\usepackage[utf8]{inputenc}
\usepackage{url}
\usepackage{amsmath}
\usepackage{mathtools}
\usepackage{bbold}
\usepackage{mathrsfs}
\usepackage[charter]{mathdesign}

\DeclareFontFamily{U}{dutchcal}{\skewchar\font=45 }
\DeclareFontShape{U}{dutchcal}{m}{n}{<-> s*[1.0] dutchcal-r}{}
\DeclareFontShape{U}{dutchcal}{b}{n}{<-> s*[1.0] dutchcal-b}{}
\DeclareMathAlphabet{\mathlcal}{U}{dutchcal}{m}{n}

\usepackage{graphicx}
\usepackage{epstopdf}
\usepackage{wrapfig}
\usepackage{braket}
\usepackage{xcolor}
\usepackage[pdfstartview=FitH,colorlinks=true,citecolor=blue,linkcolor=blue,urlcolor=blue]{hyperref}
\usepackage{float}
\usepackage{mathptmx}
\usepackage{tikz}
\usepackage[vcentermath]{youngtab}
\usepackage{cancel}

\usepackage{soul}

\begin{document}

\author{Gianni Aupetit-Diallo}
\affiliation{Université Côte d’Azur, CNRS, Institut de Physique de Nice, 06200 Nice, France}
\author{Silvia Musolino}
\affiliation{Université Côte d’Azur, CNRS, Institut de Physique de Nice, 06200 Nice, France}
\author{Mathias Albert} 
\affiliation{Université Côte d’Azur, CNRS, Institut de Physique de Nice, 06200 Nice, France}
\author{Patrizia Vignolo}
\affiliation{Université Côte d’Azur, CNRS, Institut de Physique de Nice, 06200 Nice, France}

\title{High-momentum oscillating tails of strongly interacting one-dimensional gases in a box}
\begin{abstract}
We study the equilibrium momentum distribution of strongly interacting one-dimensional mixtures of particles at zero temperature in a box potential. We find 
that the magnitude of the $1/k^4$ tail of the momentum distribution is not only due to short-distance correlations, but also to the presence of the rigid walls, breaking  the Tan relation relating this quantity to the adiabatic derivative of the energy with respect to the inverse of the interaction strength.
The additional contribution is a finite-size effect that includes a $k$-independent and an oscillating part. This latter, surprisingly, encodes information on long-range spin correlations.

 \end{abstract}

\maketitle
\section{Introduction}
\label{sec:intro}

One-dimensional (1D) quantum systems of particles with contact interactions have been the playground for theoreticians for many years since they are exactly solvable with techniques such as Bethe ansatz~\cite{LiebI,LiebII,Yang1967,Mcguire1964,Calabrese2007}
and fermionization~\cite{Girardeau1960,minguzzi2022strongly}. Access to exact solutions was and is essential to improving our understanding of the role of quantum correlations in low dimensions~\cite{Giamarchi_book,giamarchi2011}. 
During the past decades, after being realized experimentally using different particle species, trapping geometries, and adjustable interactions \cite{Mistakidis2022,minguzzi2022strongly}, the status of such systems has changed considerably.
They have gone from being toy models to one of the paradigms for quantum
 simulators \cite{Gross2017}. In turn, they can even be considered as benchmarks for other, more complex, quantum simulators \cite{Shafer2020}.
Among other examples, it is now possible to synthesize systems such as the Tonks-Girardeau (TG) gas  of strongly interacting bosons~\cite{Kinoshita1125,Parendes2004} or fermionic mixtures of $\kappa$ components with $SU(\kappa)$ interaction symmetry~\cite{pagano_one-dimensional_2014}. The gas enters the TG regime when the ratio of the interaction energy to kinetic energy becomes very large and the probability of observing two particles in the same position becomes approximately zero.

Due to the diluteness of ultracold gases, atomic interactions  can be well approximated by a zero-range  potential and an important consequence of strong contact interactions is the universality of many equilibrium and thermodynamic quantities, most of them being summarized by the Tan relations \cite{tan_energetics_2008,tan_generalized_2008,tan_large_2008,BARTH20112544,Patu2017}. In one of them, the interplay between contact interactions and exchange symmetry between $N$ particles leads to the appearance of a universal algebraic behavior of the tail of the momentum distribution of the form $\mathcal{K}_N/k^4$ for momentum $\hbar k$ larger than any other typical momentum scale, such as the Fermi momentum $k_\mathrm{F}$.

$\mathcal{K}_N$ is usually identified with $\mathcal{C}_N$, Tan's contact, which is proportional to  $\partial E/\partial g^{-1}$, namely to $gE_\mathrm{int}$, the product between the interaction strength and the total interaction energy of the system \cite{lenard_momentum_1964,Minguzzi02,olshanii_short-distance_2003}. The equivalence $\mathcal{K_N}=\mathcal{C}_N$ holds at equilibrium for both homogeneous systems with periodic-boundary conditions and smoothly trapped systems, for any mixture of interacting particles, and any dimension \cite{tan_energetics_2008,tan_generalized_2008,tan_large_2008,BARTH20112544}. 

The origin of the $1/k^4$ decay is the universal way the many-body wavefunction has to accommodate the contact interaction when two particles approach each other. 
For instance, antisymmetric exchanges neutralize the effects of contact interactions and do not contribute to $\mathcal{K}_N$,
while symmetric exchanges induce in the many-body wavefunction, and thus in the one-body reduced density matrix (OBDM), a discontinuity of the derivative, a cusp, that contributes to the $\sim 1/k^4$ behavior of the momentum distribution tail \cite{Minguzzi02,olshanii_short-distance_2003,vignolo2013}. $\mathcal{K}_N$ is therefore sensitive to the exchange symmetry and can be used as observable for symmetry spectroscopy in quantum mixtures \cite{decamp_high-momentum_2016,decamp_strongly_2017}. This interplay of contact interactions and symmetry has repercussions on the spectrum of the finite interaction system \cite{volosniev_strongly_2014,decamp_high-momentum_2016}. 

However, violations of the Tan relation have been pointed out in nonequilibrium scenarios, induced by impurities~\cite{Cayla2023}, particle losses~\cite{PhysRevLett.126.160603}, interaction quenches~\cite{corson2016, rylands2023}, three-body effects~\cite{colussi2020}, and at equilibrium for high temperatures~\cite{derosi2023}. In this Letter, we show how the presence of a box confining potential also breaks down the Tan relation for 1D gases at equilibrium and zero temperature.  We find that  $\mathcal{K}_N$ not only has an average value larger than $\mathcal{C}_N$, but also, for strong interactions, develops oscillations [cf. Fig.~\ref{fig1}],  which are connected to the spin-coherence properties of the gas from one border of the box to the other.

Trapping atoms in optical-box potentials is becoming increasingly popular over the last years, and has led to important results in three-dimensional and two-dimensional gases~\cite{navon2021box}. Therefore, this work aims to guide future experiments using box potentials in one dimension.

We exemplify our findings using two canonical systems as spinless non interacting fermions and TG bosons (Sec.~\ref{sec:one-comp}), before generalizing them to arbitrary mixtures of quantum particles with infinite interactions (Sec.~\ref{part 4}).

\section{General considerations}
\label{sec:gen_cons}
In the following, we briefly recall the definition of the momentum distribution and explain how its large-momenta tail is usually related to short-distance correlations in the system, thus to Tan's contact~\cite{olshanii_short-distance_2003,BARTH20112544, WernerCastin_ferm, WernerCastin_bos}. The momentum distribution $n(k)$, that is, the average density of particles with momentum $\hbar k$, can be expressed as the Fourier transform of the OBDM $\rho_1(x, y)$:
\begin{eqnarray}
  &n(k)=\dfrac{1}{2\pi}\int\limits_{\mathcal{D}^2} dx\, dy\, \rho_1(x,y) e^{-i k(y-x)},\label{eq:nk_def}&\\
 & \!\!\!\!\rho_1(x,y)\!=\!N\!\!\!\!\!\!\int\limits_{\mathcal{D}^{N-1}}\!\!\!\!\!\!  dx_2 \cdots dx_N \,\Psi^* (x, x_2,...,x_N) \Psi(y, x_2,...,x_N),\label{rho1}&
\end{eqnarray}
 where $\Psi(x_1, x_2,...,x_N)$ is the many-body wave function of $N$ particles, which, at this stage, can describe any kind of mixture. The integration domain $\mathcal{D}$ runs over the entire system  and depends on the considered geometry.
 
In this Letter, we focus on the large-momenta tail of $n(k)$ given by 
\begin{equation}
n(k) \underset{k\rightarrow\infty}{\simeq} \frac{\mathcal{K}_N}{k^4},
  \label{C}
\end{equation}
where the power-law decay derives from the type of singularity of $\rho_1(x, y)$ or $\Psi(x_1, \dots, x_N)$ and its weight $\mathcal{K}_N$ depends on the function slope in the vicinity of these singularities and on their number.

The origin of  the $1/k^4$ tail can be understood by mathematical means. According to Watson's lemma~\cite{bleistein_asymptotic_1986, olshanii_short-distance_2003}, the asymptotics of the Fourier transform of functions  which have a singularity of the type $f(z)=F(z)|z-z_0|^\alpha$, with $F(z)$ analytic, and $\alpha > -1$ and $\alpha \neq 0, 2, 4, \dots$ reads

\begin{equation}
  \int\limits_{\mathcal{D}} dz~e^{-ikz}F(z)|z-z_0|^\alpha\!\!\underset{k\rightarrow\infty}{=}\!\!\mathcal F_\alpha \frac{e^{-ikz_0} F(z_0)}{|k|^{\alpha+1}}+\mathcal{O}\left(\frac{1}{|k|^{\alpha+2}}\right), 
    \label{asymptTF}
\end{equation}
where $ \mathcal F_\alpha= 2\cos[\pi(\alpha+1)/2]\Gamma(\alpha+1)$ and $\Gamma(\alpha)$ is the Gamma function. Therefore, by looking at Eqs.~\eqref{eq:nk_def} and~\eqref{rho1}, the possible contributions to the $1/k^4$ tail of $n(k)$ could be seen as non-analytic terms of the form: (i) $|x-y|^3$ in $\rho_1(x,y)$~\cite{forrester_finite_2003, vignolo2013} or (ii)  $|x-\bar x|$, with $\bar x \in \mathcal{D}$, in $\Psi(x, x_2, ..)$~\cite{Minguzzi02,olshanii_short-distance_2003}.

A pedagogical example of this behavior is provided by the TG gas for a smooth trapping potential. Its OBDM behaves as $|x-y|^3$ around $x\sim y$ and, consequently, $n(k)$ displays an algebraic tail~\cite{forrester_finite_2003}. This differs from the case of free fermions and  bosons in the same trap configuration, whose OBDMs are instead analytical in $\mathcal{D}$ and their momentum distributions do not have any algebraic tail~\cite{note1}. This can be shown by expanding the OBDM for the TG gas, $\rho_1^{TG}$,  in terms of the spinless fermions reduced density matrices  $\rho_{1+j}^{F}$ with $j\geq 1$ \cite{lenard_onedimensional_1966, forrester_painleve_2003}, namely,
\begin{equation}
    \begin{split}
      \rho_1^\mathrm{TG}(x,y)&=\rho_1^\mathrm{F}(x,y)+\sum_{j=1}^{N-1}\dfrac{(-2)^j}{j!} \int_x^{y}\!\!dx_{2}\dots dx_{1+j}\\&\times\,\rho_{1+j}^\mathrm{F}(x,x_2,\dots;y,x_2,\dots),
    \end{split}
    \label{lenseriesmain1}
\end{equation}
for $x<y$. For a smooth trapping potential, the only term that contributes to the $1/k^{4}$ algebraic
decay of $n(k)$ is the first term of the expansion in Eq.~(\ref{lenseriesmain1}), namely, $-2\int_x^{y}\!\!dx_{2}\rho_{2}^{F}(x,x_2;y,x_2)$ \cite{olshanii_short-distance_2003}. Indeed, by using  Eq.~\eqref{lenseriesmain1} and introducing the change of coordinates $x_r=y-x$ and $X=(x+y)/2$, one has~\cite{Fang09,vignolo2013}

\begin{equation}
\begin{split}
   n^\mathrm{TG}(k)&\underset{k\rightarrow\infty}{\simeq}\dfrac{1}{2}\int\limits_{2\mathcal{D}} dx_{r}\, e^{-i kx_r} \dfrac{|x_r|^3}{6} \mathcal{C}_{N}^\mathrm{TG},
 \end{split} 
 \label{eq:nkTG_ring}
\end{equation}
where $1/k^4$ is given by applying Eq.~\eqref{asymptTF} to the integral in $x_r$ and 
\begin{equation}
 \mathcal{C}_{N}^\mathrm{TG} \equiv \frac{2}{\pi}   \int\limits_{\mathcal{D}}dX\,\lim_{\varepsilon\rightarrow 0}\dfrac{\rho_{2}^\mathrm{F}(X-\varepsilon,X;X+\varepsilon,X)}
    {\varepsilon^2}
    \label{c-rho2}
\end{equation}
is the Tan contact, which is equivalent to $\mathcal{K}_N$ in this case. Equation~(\ref{c-rho2}) enlightens the role of two-body correlations in $\mathcal{C}_{N}^{TG}$.

In the next sections, we will show how the presence of a box potential adds to the bulk term $\mathcal{C}_{N}^\mathrm{TG}$ an edge contribution. This is not only a trivial consequence of the cancellation of the wave function at the border, but also an interplay between rigid-border effects and coherence properties of the gas.

\section{One-component gases in a box}
\label{sec:one-comp}

In order to investigate the effect of hard walls, we will discuss in this section two simple examples of 1D quantum gases at equilibrium in a box geometry ($x\in[-L/2,L/2]$) and at zero temperature~\cite{note2}.  We begin with a non-interacting Fermi gas whose momentum distribution does not have any algebraic tail in the homogeneous ring trap as well as in the presence of smoothly varying potentials. 
The second example will be the TG gas in a box, whose many-body wave function only differs from the one of spinless fermions by the particle-exchange symmetry. 

\begin{figure}
    \centering
    \includegraphics[scale=0.6]{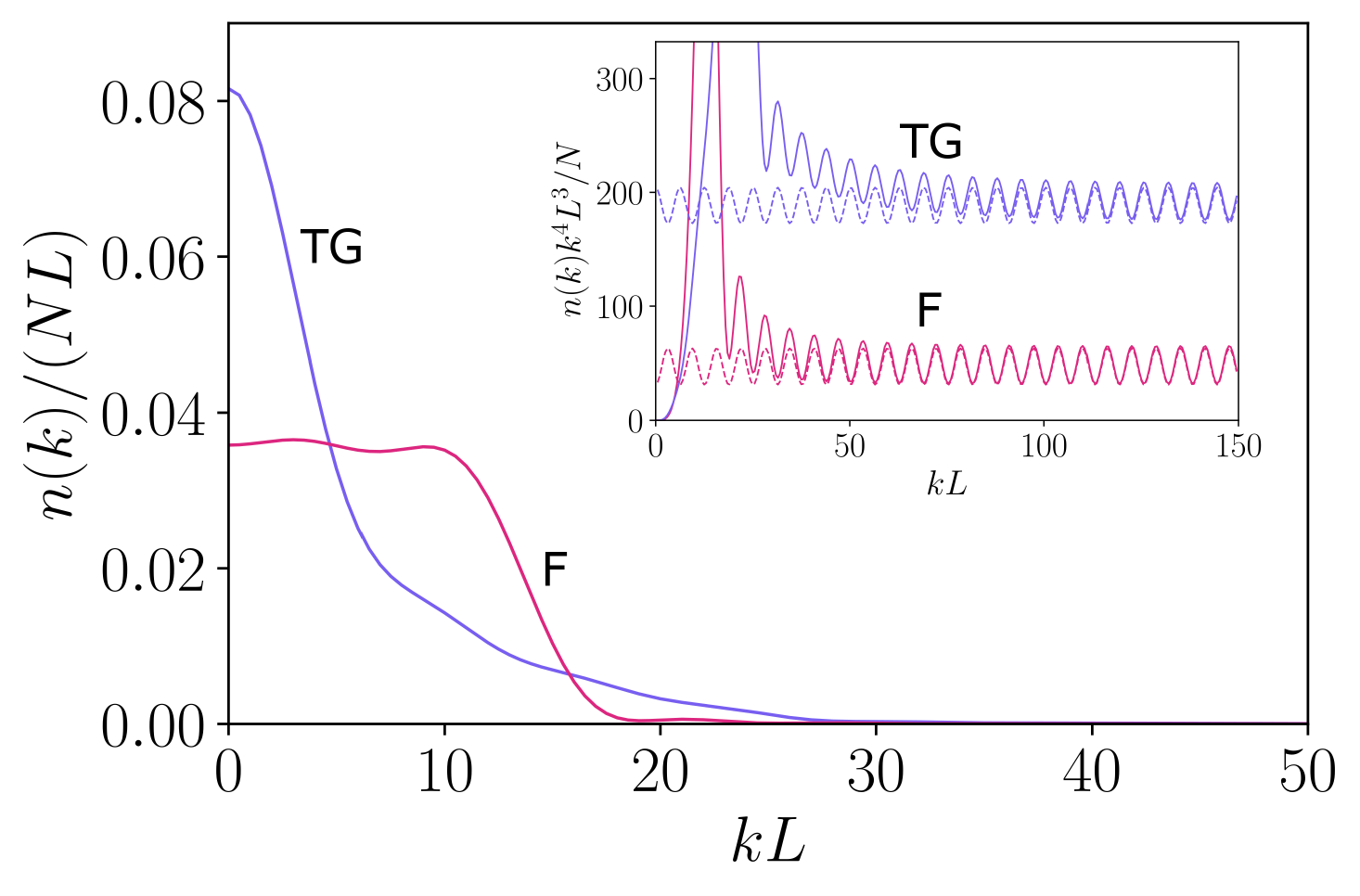}
    \caption{Normalized momentum distribution $n(k)/N$, in units of $1/L$, as a function of $kL$ for four spinless fermions (magenta) and four TG bosons (light violet).
    In the inset, the solid lines are the same $n(k)/N$ multiplied by $k^4$, in units of $L^3$, while the dashed lines correspond to the asymptotic analytical expressions given in Eqs. (\ref{eq-f}) and (\ref{CTG}).}

    \label{fig1}
\end{figure}

\subsection{Spinless fermions}
\label{sec:spinless}
We now consider 1D spinless fermions trapped in a box of size $L$. In this case, the many-body wave function is simply the Slater determinant of $N$ lowest energy single-particle orbitals and the OBDM takes the form~\cite{meckes_random_2019,Lacroix-a-chez-toine}
\begin{equation}
\begin{split}
    \rho_1^\mathrm{F}(x,y)&=\dfrac{1}{2L}\left[\dfrac{\sin\left[(2N+1)\frac{\pi}{2L}(x-y)\right]}{\sin[\frac{\pi}{2L}(x-y)]}\right.\\
    &\left.-\dfrac{\sin[(2N+1)\frac{\pi}{2L}(x+y+L)]}{\sin[\frac{\pi}{2L}(x+y+L)]}\right],
    \end{split}
\end{equation}
with $|x|,|y|\le L/2 $.
As explained in Sec.~\ref{sec:gen_cons}, the calculation of the momentum distribution tail boils down to the investigation of nonanalyticities in the OBDM. In this case, they are only located at the edges, and the momentum distribution of the spinless Fermi gas develops an algebraic oscillating tail such that~\cite{Bruyne2021,suppl}
\begin{equation}
\mathcal{K}^\mathrm{F}_N =\mathcal{B}_N+(-1)^{N+1}\mathcal{A}_N\cos\left(kL\right),
\label{eq-f}
\end{equation}
with $\mathcal{A}_N=N(N+1)\pi/L^3$ 
and $\mathcal{B}_N=(2N+1)\mathcal{A}_N/3$.
 The $k$-independent part $ \mathcal{B}_N$ comes from contributions where $x$ and $y$ are close to the same edge. Roughly speaking, the effect of a hard wall in $L/2$ ($-L/2$) introduces a half cusp, with respect to the coordinates $x$ and $y$, of the form $|x- L/2|$ and $|y- L/2|$ ($|x+ L/2|$ and $|y+ L/2|$). Instead, the oscillating part is given by contributions of half cusps at opposite walls $(x\rightarrow\pm L/2,y\rightarrow\mp L/2)$. At first sight, it could be seen as an effect of diffraction by the box. However, its interpretation is more subtle and will become clearer when we will consider the case of a general mixture in Sec. \ref{part 4}. To support our conclusions, we have computed numerically the momentum distribution of a spinless Fermi gas of $N=4$ particles, and we have compared it with the asymptotic behavior given in Eq. (\ref{eq-f})  (see Fig. \ref{fig1}). 

\subsection{Tonks-Girardeau bosons} 
 In order to calculate the asymptotic behavior of the momentum distribution for $N$ TG bosons trapped in a box, we start from the OBDM expressed as an expansion in terms of the spinless fermions $n$-body density matrices, as shown in Eq. (\ref{lenseriesmain1}).
In the presence of smooth trapping potentials, we have already seen that only the first term of the series contributes to the contact.
For the TG in a box, we can individuate three different  contributions to the $1/k^{4}$ tail of the momentum distribution. The first contribution comes 
from $\rho_1^\mathrm{F}(x,y)$ and gives the terms in Eq. (\ref{eq-f}). This contribution is similar to the result found in Ref.~\cite{PhysRevLett.126.160603} showing that the discrepancy between $\mathcal{K}_N$ and $\mathcal{C}_N$ in a Lieb-Liniger gas with losses is due to the contribution of the rapidities.
The second contribution comes from $-2\int_x^{y}\!\!dx_{2}\rho_{2}^\mathrm{F}(x,x_2;y,x_2)$ and gives the usual Tan contact $\mathcal{C}_N$ [Eq. (\ref{c-rho2})], connected to the short-distance two-body correlations. 
For $N$ TG bosons in a box, we obtain
\begin{equation}
    \mathcal{C}_N^\mathrm{TG}=\dfrac{N(N^2-1)(2N+1)}{3L^3}\pi=(N-1)\mathcal{B}_{N}.
\end{equation}
Indeed, the two half cusps in $(+L/2,+L/2)$ and $(-L/2,-L/2)$ contributing to $\mathcal{B}_{N}$ have the same weight and scaling as the $(N-1)$ interparticles TG cusps of the bulk contribution $\mathcal{C}_N^{TG}$.

Remarkably, there is a third, nonlocal contribution entering the momentum distribution tail which can be derived by  integrating {\it all} the higher-order fermionic density matrices of the second term in Eq. (\ref{lenseriesmain1}) over {\it all} the system. Indeed, it can be shown that~\cite{suppl}
 \begin{equation}
\begin{split}
& \lim_{\substack{x\rightarrow -\frac{L}{2}\\ y\rightarrow \frac{L}{2}}}\sum_{j=1}^{N-1}\dfrac{(-2)^j}{j!}
         \prod_{\ell=2}^{j+1}\int_x^{y} dx_{\ell}\,\rho_{1+j}^\mathrm{F}(x,x_2,\dots;y,x_2,\dots) \\
        & = -2 \rho_{1}^\mathrm{F}(x,y)|_{x\sim -\frac{L}{2}, y\sim \frac{L}{2}},
         \end{split}
         \label{pat-gian}
\end{equation}
if $N$ is even and 0 otherwise. Such a term changes the sign of the oscillating part, with respect to the fermionic case {\it if the number of particles is even}. This means that for the TG gas, the sign of the oscillating part does not depend on the number of trapped bosons. Ultimately, we find that the asymptotic behavior of the momentum distribution for $N$ TG bosons in the box can be written as
  \begin{equation}
  \begin{split}
    \mathcal{K}^\mathrm{TG}_N&=  \mathcal{C}_{N}^\mathrm{TG}
    +\mathcal{B}_N+ \mathcal{A}_{N}\cos\left(kL\right)\\
    &=\dfrac{N}{N-1}\mathcal{C}_{N}^\mathrm{TG}+ \mathcal{A}_{N}\cos\left(kL\right).
    \end{split}
    \label{CTG}
 \end{equation}
The average effect of the border ($\mathcal{B}_N$) is equivalent to the addition of a boson to the system. Moreover, it induces oscillations of the same amplitude as for a spinless Fermi gas, but with a phase that does not depend on the particle number parity. In order to elucidate this result, we plot in the inset of Fig. \ref{fig1} the comparison between Eq. (\ref{CTG}) and the numerical calculation of $\mathcal{K}_N$ for the case of $N=4$ particles. 
Notice that, in the thermodynamic limit, we recover the known result for the contact density $\mathcal{C}_N^\mathrm{TG}/L$ of a homogeneous TG gas with density $\mathlcal{n}=N/L$:
$\lim_{N,L\rightarrow\infty}\mathcal{K}_N^{\mathrm{TG}}/L= \lim_{N,L\rightarrow\infty}\mathcal{C}_N^\mathrm{TG}/L=\frac{2}{3}\mathlcal{n}^4\pi$ \cite{Decamp2018}.

\section{Mixtures in a box} \label{part 4}
 We now generalize our results to strongly interacting bosonic and/or fermionic mixtures.

\subsection{Tonks-Girardeau limit for mixtures}
We consider a 1D mixture of $N$ particles with $\kappa$ components and interacting via a two-body contact interaction. The Hamiltonian for this system is given by
 \begin{equation}
 \begin{split}
 \hat{H}&=\sum_{\sigma\sigma'}^{\kappa}\sum_i^{N_\sigma}
 \left[-\frac{\hbar^2}{2m}\frac{\partial^2}{\partial x_{i,\sigma}^2}+g_{\sigma\sigma'}\sum_{j>i}^{N_{\sigma'}} \delta(x_{i,\sigma}-x_{j,\sigma'})\right],
 \end{split}
 \label{ham}
 \end{equation}
 where $i,j \in [1, N]$ and $\sigma,\sigma' \in [1, \kappa]$ are the particle and spin indices, respectively, and $g_{\sigma,\sigma'}$ is the inter- ($\sigma\neq\sigma'$) or intra-species ($\sigma=\sigma'$) interaction. Remarkably, the latter one is zero for identical fermions interacting via $s$-wave contact interactions.  In the limit $g_{\sigma\sigma'}\rightarrow +\infty$, for any $\sigma,\sigma'$, the many-body wave function $\Psi$ vanishes whenever $x_i=x_j$. Thus, $\Psi$ can be written as follows~\cite{volosniev_strongly_2014, deuretzbacher_quantum_2014}:
 \begin{equation}
\Psi(X)=\sum_{P\in S_N}a_P\theta_P(X)\Psi_A(X),
\label{vol}
\end{equation}
where $X=(x_{1,\sigma_1},\dots,x_{N,\sigma_N})$ collects particle and spin indices, the index $P$ indicates a permutation inside the permutation group of $N$ elements, $S_N$, $\theta_P(X)$ is the generalized Heaviside function, which is
equal to 1 in the coordinate sector $x_{P(1),\sigma_{P(1)}}<\dots<x_{P(N),\sigma_{P(N)}}$ and 0 elsewhere, and $\Psi_A$ is the wave function for $N$ spinless fermions. In particular, $\Psi_A$ is the Slater determinant built from the natural one-particle orbitals of the box.

Because of the statistics of identical particles, we can restrict the sum over $P$ in Eq.~\eqref{vol} to $N!/\prod_\sigma N_\sigma!$
independent elements instead of $N!$.
These groups of sectors represent all the possible spin configurations and are usually called snippets \cite{Fang2011,volosniev_strongly_2014}. They constitute the proper basis for describing a multicomponent spin mixture and will be used throughout this section.

Moreover, in the strongly interacting limit,
both for (i) the SU($\kappa$) case  
$1/g_{\sigma\sigma}=1/g_{\sigma\sigma'}=1/g\ll 1$
and (ii) the broken symmetry one $1/g_{\sigma\sigma'}=1/g\ll 1$ with $1/g_{\sigma\sigma}=0$~\cite{volosniev_engineering_2015,Aupetit2022}, the Hamiltonian in Eq.~\eqref{ham} can be mapped into the spin Hamiltonian
\begin{equation}
    \hat{H}\underset{g\rightarrow\infty}{\rightarrow} E_F\mathbb{1}+H_\mathrm{eff}(J_N),
    \label{eq:Heff}
\end{equation}
where $E_F$ is the Fermi energy related to the noninteracting system and  $H_\mathrm{eff}=J_N(- (N-1)\mathbb{1}\pm\sum_j\hat{P}_{j,j+1})$,  $\hat{P}_{j,j+1}$ being the permutation operator exchanging two interacting next-neighboring particles, and 
$J_N=(\hbar^4/m^2)\alpha_N/g$ is the coupling constant with $\alpha_N=2E_F m/(\hbar^2L)$ $=\mathcal{B}_N\pi/2$  the nearest-neighbor exchange term~\cite{deuretzbacher_momentum_2016}.
    Indeed, in the homogeneous system, $\hbar^2L \alpha_N/m$ is twice the total kinetic energy, which is connected to the slope of the cusps \cite{Barfknecht2021,Aupetit2022}.

For a multicomponent system, the OBDM can be written as 
$\rho_1(x,y)=\sum_\sigma N_\sigma \rho_{1,\sigma}(x,y)$
    with
    \begin{equation}
    \rho_{1,\sigma}(x,y)=\sum_{i,j=1}^N c^{(i,j)}_{\sigma}\rho^{(i,j)}(x,y),
    \label{g_1}
\end{equation}
where  $\rho^{(i,j)}(x,y)$ and $c^{(i,j)}_{\sigma}$ are the spatial and spin parts calculated on the sector $x_{1,\sigma_1}<\dots<x_{i-1,\sigma_{i-1}}<x<x_{i+1,\sigma_{i+1}}<\dots<x_{j,\sigma_j}<y<x_{j+1,\sigma_{j+1}}<\dots<x_{N,\sigma_N}$~\cite{deuretzbacher_momentum_2016}. In particular,
\begin{equation}
    c_\sigma^{(i,j)}=\delta^{\sigma}_{\sigma_i} \sum_{P\in S_N} a_{P} a_{P_{i\rightarrow j}}, 
    \label{eq:cij}
\end{equation}
where $\delta^{\sigma}_{\sigma_i}$ selects only the sites with spin $\sigma_i=\sigma$ and $a_{P_{i\rightarrow j}}$ is the sector coefficient obtained by starting from the spin configuration labeled as $a_P$ and applying a  cyclic permutation which takes the $i$-th element into the $j$-th position, and vice versa.

\begin{figure}
    \centering
    \includegraphics[scale=0.6]{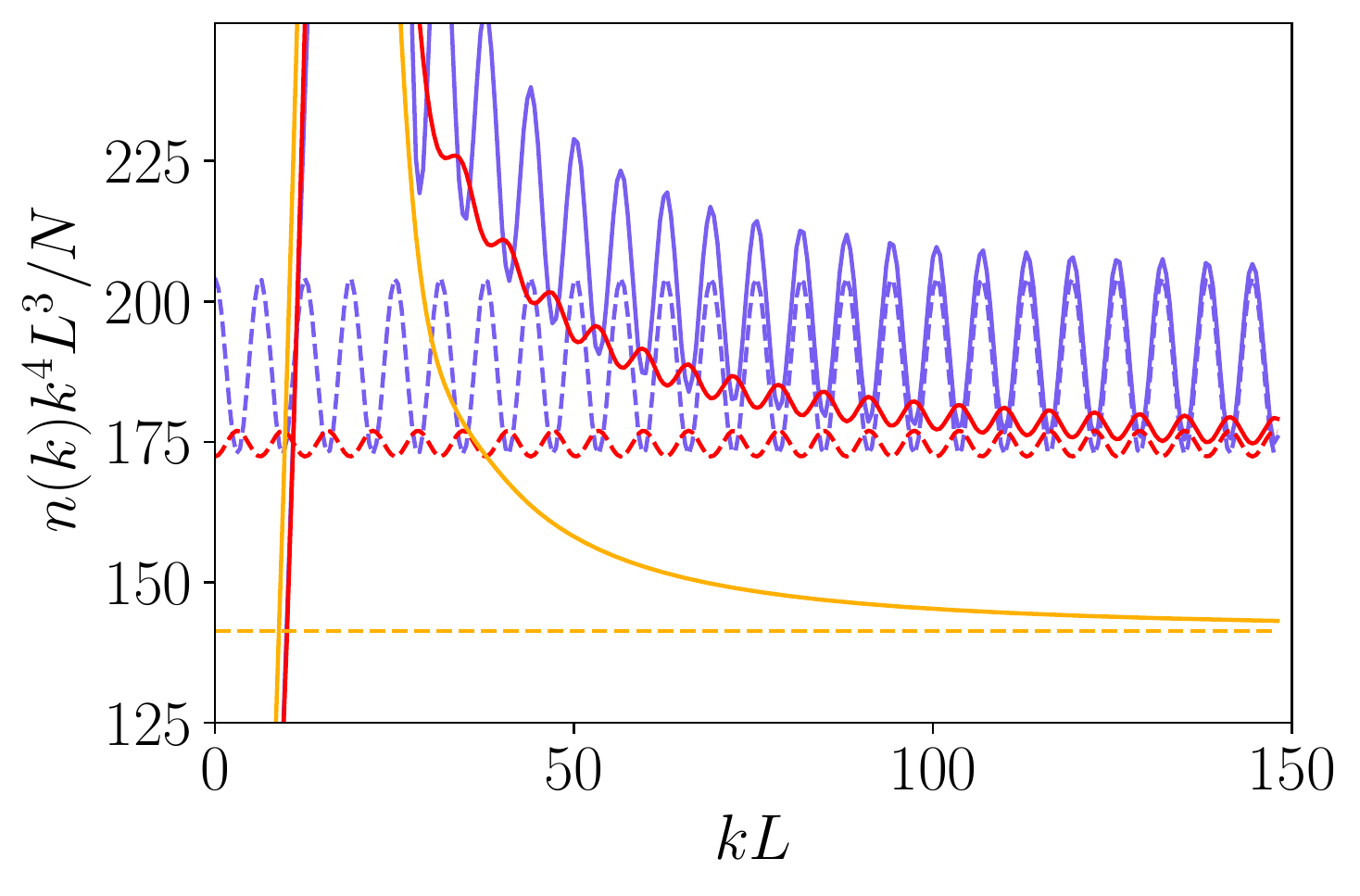}
    \caption{The solid lines stand for normalized momentum distribution $n(k)/N$ multiplied by $k^4$, in units of $L^3$, for the case of $2+2$ SU(2) bosons (i) in the ground state (violet, upper curve), (ii) in the first excited state (orange, central curve), and (ii) in the third excited state (yellow, lower curve). 
    The dashed lines stand for the analytical expression of $\mathcal{K}_N^\mathrm{mix}/N$, Eq. (\ref{CMix-separate}), evaluated for cases
    (i), (ii), and (iii) (same color code).}
    \label{fig2}
\end{figure}
\subsection{$\mathcal{K}_N$ for a mixture in the TG limit}
All the elements required to compute the generalization of Eqs.~(\ref{eq-f}) and~(\ref{CTG}) have been presented. For mixtures, the development is similar and is detailed in Ref.~\cite{suppl}. The asymptotic behavior of the momentum distribution in the case of spin mixtures assumes the form
\begin{equation}
    \begin{split}
    \mathcal{K}_N^\mathrm{mix}=&\mathcal{C}^\mathrm{mix}_{N}\!+\!\mathcal{B}_{N}\!+\!(-1)^{N\!+\!1}\!\!\mathcal{A}_{N}\sum_{\sigma}\frac{N_\sigma}{N} c_{\sigma}^{(1,N)}\!\!\cos(kL)\\
    =&\dfrac{S\!+\!1}{N\!-\!1}\mathcal{C}_N^\mathrm{TG}\! +\!(-1)^{N\!+\!1}\!\!\mathcal{A}_{N}\sum_{\sigma}\frac{N_\sigma}{N} c_{\sigma}^{(1,N)}\!\!\cos\left(kL\right).
    \end{split}
    \label{CMix-separate}
\end{equation}
The quantity
\begin{equation}
  S=\sum_{P}\sum_{i=1}^{N-1}\left[\frac{1}{4}(a_P-a_{P_{i,i+1}})^2(1- \delta^{\sigma_{i+1}}_{\sigma_i})+\eta a_{P}a_{P_{i, i+1}}\delta^{\sigma_{i+1}}_{\sigma_i}\right]
  \label{eq:S}
\end{equation}
takes into account the number of symmetric exchanges between particles~\cite{Aupetit2022} and is proportional to the eigenvalue of the rescaled effective Hamiltonian $H'_\mathrm{eff}=H_\mathrm{eff}/J_N$ [see Eq.~(\ref{eq:Heff})]. $P$ runs over the snippets and $\eta$
is equal to 1 for identical bosons and 0 otherwise. As expected, we can recover Eqs. (\ref{eq-f}) and (\ref{CTG}) for the cases of spinless fermions and TG bosons, respectively. Indeed, it can be shown~\cite{suppl} that for spinless fermions $S=0$ and $c_\sigma^{(1,N)}=1$,  and for a TG gas $S=N-1$ and  $\sum_{\sigma}\frac{N_\sigma}{N}c_\sigma^{(1,N)}=(-1)^{N+1}$ .

As for the one-component cases, the large-$k$ tail of $n(k)$ is not given solely by the Tan contact, but includes two additional terms. The first, the $k$-independent contribution $\mathcal{B}_N$, does not depend on the type of particles or mixture and counts such as an extra symmetric exchange in the mixture. The second, the oscillating part, is more intriguing, since the amplitude of the oscillations depends, remarkably, on long-distance spin correlations. Indeed, the only term of Eq.~\eqref{g_1} that does not vanish in the limit $x\to -L/2$ and $y\rightarrow+L/2$ corresponds to the cyclic permutation $P_{1\to N}$~\cite{suppl}.  Therefore,  $c_\sigma^{(1,N)}$ can be interpreted as the one-body spin correlation through the whole system.

For the case of SU($\kappa$) bosonic or fermionic mixtures, the oscillation amplitude is maximal when the spin correlation is maximal, that is when the state of the system is equivalent to a single-component gas~\cite{Aupetit2022} (meaning the ground state for bosonic and the most excited state for fermionic mixtures). Contrarily, as shown in Fig. \ref{fig2}, it vanishes for some particular cases where long-distance spin correlation is absent~\cite{suppl}.
The oscillation phase fluctuates by a factor $\pi$ depending, among other things, on the number of particles, for almost all states, except for the ground state of a SU($\kappa$) bosonic mixture~\cite{suppl}.

\section{Concluding remarks}
In conclusion, we have shown that the presence of a hard wall trapping potential breaks down the Tan relation connecting the $1/k^4$ decay of the momentum distribution of a 1D gas characterized by repulsive contact interactions to the adiabatic derivative of the energy with respect to the inverse of the interaction strength, even if the system is at equilibrium. In the strongly interacting limit, the presence of the two hard walls has a double effect.
The first is rather trivial: it mimics the presence of an additional boson or impurity in the system.
The second is more subtle: the tails develop oscillations whose {\it amplitude depends on the nonlocal spin correlations over the whole system size}. 
The sign of this contribution depends generally on the number of
particles, except for the ground state of a bosonic $SU(\kappa)$ mixture. This can be of particular interest for experiments. In ultracold gases, the momentum distribution can be measured by switching off the trapping potential and imaging the cloud after a ballistic expansion~\cite{Parendes2004, pagano_one-dimensional_2014}, and these measurements are obtained by averaging over system realizations where the number of particles fluctuates shot to shot. Therefore, the observation of oscillating tails in $n(k)$ of $SU(\kappa)$ bosonic mixtures could be used to determine whether the system is mainly cooled down in its ground state or not. Indeed, only in the first case the oscillation amplitude will be not vanishing. Let us remark that identifying the exact populated state might be experimentally difficult since the spectrum of a strongly interacting mixture is characterized by the presence of a large number of states very close in energy to the ground state~\cite{deuretzbacher_exact_2008,decamp_high-momentum_2016}.

Finally, our study can be extended to finite temperatures and different dimensions, and out-of-equilibrium scenarios, such as spin-mixing  dynamics~\cite{pecci2022}.

\section*{Acknowledgments}
We would like to thank Anna Minguzzi for useful discussions.
 We acknowledge funding from the ANR-21-CE47-0009 Quantum-SOPHA project. P.V. is a member of the Institut Universitaire de France.


\begin{thebibliography}{54}
\bibitem{note1} For example, in a homogeneous trap ring, the momentum distribution for free fermions is a step Fermi function and for free bosons is a Dirac $\delta(k)$.
\bibitem{note2} Here, we consider an ideal box, but in the experiments, the walls of the trap could have a local curvature $b$. In that case, our results would be valid for $k < 1/b$.
\end{thebibliography}
\end{document}


\title{Supplemental material for\\ ``High-momentum oscillating tails of strongly interacting one-dimensional gases in a box''}

\author{Gianni Aupetit-Diallo}
\affiliation{Université Côte d’Azur, CNRS, Institut de Physique de Nice, 06200 Nice, France}
\author{Silvia Musolino}
\affiliation{Université Côte d’Azur, CNRS, Institut de Physique de Nice, 06200 Nice, France}
\author{Mathias Albert}
\affiliation{Université Côte d’Azur, CNRS, Institut de Physique de Nice, 06200 Nice, France}
\author{Patrizia Vignolo}
\affiliation{Université Côte d’Azur, CNRS, Institut de Physique de Nice, 06200 Nice, France}

\maketitle 
\section{Two-body wave function in a box}
\label{sec:2bodyWF}

\begin{figure}[tbh]
    \centering
    \includegraphics[scale=0.7]{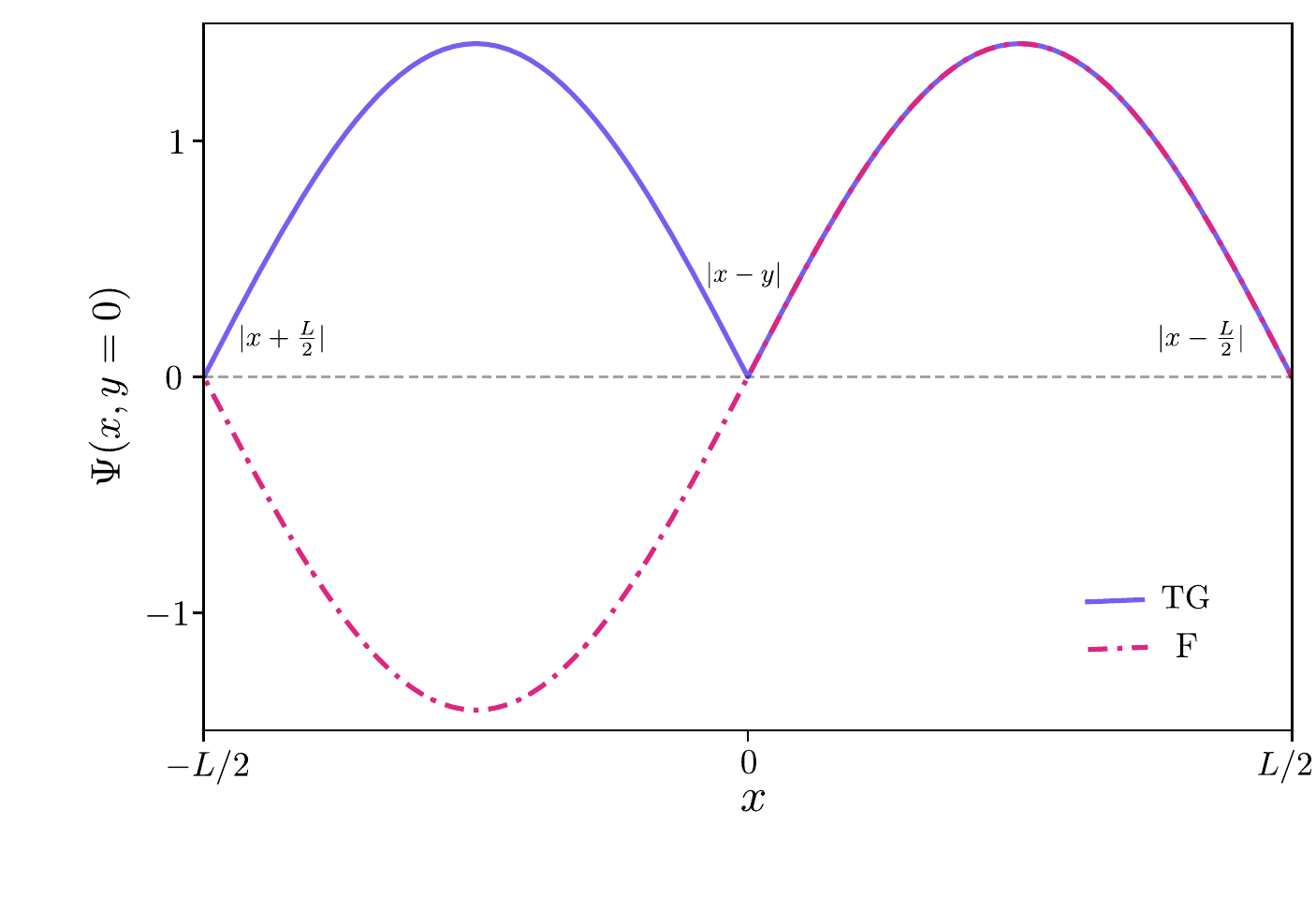}
    \caption{Two-body wave function $\Psi(x, y)$ with $y=0$ for spinless fermions (solid line) and Tonks-Girardeau bosons (dash-dotted line) trapped in a box of size $L$. We clearly distinguish three different points of non-analyticity, at $|x \pm L/2|$ (for fermions and bosons) and $|x- y|$ (only for bosons). }
    \label{fig:2bWF}
\end{figure}

To make more evident the different origins of the terms contributing to the tail of the momentum distribution $n(k)$, we discuss in this section the simple cases of  $N=2$ spinless fermions and Tonks-Girardeau bosons in a box of size $L$. As discussed in the main text, inside a box,
the power-law decay $1/k^4$ of $n(k)$ derives from the singularities of $\rho_1(x, y)$ when $x \to y$ with $x, y \in [-L/2, L/2]$ or when $x \to \pm L/2$ and $y \to \pm L/2$ or $ \mp L/2$. These cusps can be easily seen by looking at the two-body wave function $\Psi(x, y)$ for spinless fermions and Tonks-Girardeau bosons. 
Therefore, in Fig.~\ref{fig:2bWF}, we show the two-body wave function $\Psi(x, y)$ at fixed $y=0$ for the systems under consideration. We notice that, for fermions, the function is not analytic only when $x\to \pm L/2$. Instead, for Tonks-Girardeau bosons, there is an extra singular point, namely, when  $x=y=0$, due to the fact that  $\Psi^\mathrm{TG}(x, y) = |\Psi^\mathrm{F}(x, y)|$~\cite{Girardeau1960}.  The latter contributes to the Tan contact, which is nonzero only for bosons.

We remark that, contrary to the cusp in the middle of Fig.~\ref{fig:2bWF}, the singularities at the borders do not appear if we plot $\Psi(x, y)$ as a function of the relative distance $x-y$. Therefore, they cannot be derived by a change of coordinates in $\rho_1(x,y)$ as done in Eq. (6) of the main text.

\section{Proof of the large-momenta tail of the momentum distribution in the box}

In this section, we present the derivation of the large-$k$ tail of the momentum distribution for spinless fermions (Sec.~\ref{A1}), Tonks-Girardeau gases (Sec.~\ref{A2}), and mixtures (Sec.~\ref{sec:gen_cas}) (Eqs. (9), (12) and (18) of the main text, respectively).

\subsection{Spinless fermions}
 \label{A1}
 As a reminder, the large-$k$ tail of $n(k)$ for $N$ spinless fermions (Eq. (9) of the main text) is given by \cite{Bruyne2021}
\begin{align}
\lim_{k\rightarrow\infty}k^4n^\mathrm{F}(k) =\mathcal{B}_N+(-1)^{N+1}\mathcal{A}_N\cos\left(kL\right)\label{CF},
\end{align}
where $\mathcal{B}_N=(2N+1)\mathcal{A}_N/3$ is the $k$-independent part and $\mathcal{A}_N=N(N+1)\pi/L^3$ is the amplitude of the oscillating part.
We start the derivation of Eq.~\ref{CF} by writing the momentum distribution $n(k)$ for $N$ spinless fermions in a box as follows:
\begin{equation}
n(k) = \frac{1}{2\pi} \int_{-L/2}^{L/2} dx e^{ikx} \int_{-L/2}^{L/2} dy e^{-iky} \rho_1(x, y),
\label{eq:nk_spinlessF}
\end{equation}
where $\rho_1(x,y)$ is given by~\cite{meckes_random_2019,Lacroix-a-chez-toine} 
\begin{equation}
    \rho_1^\mathrm{F}(x,y)=\dfrac{1}{2L}\left[\dfrac{\sin\left[(2N+1)\frac{\pi}{2L}(x-y)\right]}{\sin[\frac{\pi}{2L}(x-y)]}-\dfrac{\sin[(2N+1)\frac{\pi}{2L}(x+y+L)]}{\sin[\frac{\pi}{2L}(x+y+L)]}\right].
    \label{eq:rho1_F}
\end{equation}
In particular,  we notice that $\rho_1^\mathrm{F}(x,x)=\rho_1^\mathrm{F}(-x,-x)$, and $\rho_1^\mathrm{F}(-x,x)=\rho_1^\mathrm{F}(x,-x)$.

We then recognize the singularities of $\rho_1$, which are in the vicinity of $|x\pm L/2|$ and $|y\pm L/2|$, and define $\mathcal{B}_N$ and $\mathcal{A}_N$ as follows
\begin{equation}
\mathcal{B}_N=\dfrac{1}{\pi}\lim_{\substack{x\rightarrow \frac{L}{2}\\ y\rightarrow \frac{L}{2}}}
    \dfrac{\rho_1^\mathrm{F}(x,y)}{|x-\frac{L}{2}||y-\frac{L}{2}|}
    =\dfrac{N(N+1)(2N+1)}{3L^3}\pi, 
    \label{pat1}
\end{equation}
and 
\begin{equation}
    (-1)^{N+1}\mathcal{A}_N=\dfrac{1}{\pi}\lim_{\substack{x\rightarrow -\frac{L}{2}\\ y\rightarrow \frac{L}{2}}}
    \dfrac{\rho_1^\mathrm{F}(x,y)}{|x+\frac{L}{2}||y-\frac{L}{2}|}=(-1)^{N+1}\dfrac{N(N+1)\pi}{L^3},
    \label{pat2}
\end{equation}
where we have divided $\rho_1(x, y)$ by its singularities to obtain a regular function and calculate the function values at those points.  Moreover, for convenience, we have included in Eqs.~\eqref{pat1} and~\eqref{pat2} the factor $1/\pi$, which comes from the $1/(2\pi)$ in the definition of the Fourier transform in Eq.~\eqref{eq:nk_spinlessF} multiplied by a factor $2$ due to the symmetry of $\rho_1(x, y)$ around the points of singularity. We remark that when one applies Watson's lemma for singularities at the edge of the domain of the definition of the wave function, one has to divide by $2$ the factor $\mathcal{F}_\alpha$ in Eq.~(4) of the main text.  Finally, the factor $(-1)^{N-1}$ in Eq.~(\ref{pat2}) is due to the change of parity of the wave function for an even (odd) number of fermions.

By looking at Eq.~\eqref{eq:nk_spinlessF}, we see that we need to apply Watson's lemma for both integrals. If we consider the case $|x\to - L/2|$ and $|y\to  L/2|$, we can write
\begin{equation}
 \int_{-L/2}^{L/2}dy e^{-iky} F(x, y) |y-L/2|\underset{k\rightarrow\infty}{\propto}e^{ikL/2}F(x, L/2)\dfrac{1}{k^2},
\label{eq:int1}
\end{equation}
and
\begin{equation}
 \int_{-L/2}^{L/2}dx e^{ikx}G(x, L/2)|x+L/2|\underset{k\rightarrow\infty}{\propto}e^{ikL/2}G(-L/2, L/2)\dfrac{1}{k^2},
\label{eq:int2}
\end{equation}
where $F$ and $G$ are analytic in the specific domain of integration. Finally, one can derive the $1/k^4$ decay by inserting Eqs.~\eqref{eq:int1} and~\eqref{eq:int2} in Eq.~\eqref{eq:nk_spinlessF}. We then repeat the same steps for $|x\to  L/2|$ and $|y\to - L/2|$, and we obtain an analogous contribution with an opposite sign in the exponential.

Finally, by summing up the contributions of the two singularities $|x\pm L/2||y\mp L/2|$, we obtain
\begin{equation}
    n(k)k^4\underset{k\rightarrow \infty}{\propto}e^{ik L} + e^{-ikL}= 2 \cos{(kL)},
    \label{eq:cos}
\end{equation}
which appears in Eq.~(\ref{CF}).
\subsection{Tonks-Girardeau bosons \label{A2}}
As a reminder, the large-$k$ tail of $n(k)$ for the Tonks-Girardeau gas (Eq. (12) of the main text) is given by

  \begin{equation}
    \lim_{k\rightarrow \infty}k^4n^\mathrm{TG}(k)= \mathcal{C}_{N}^\mathrm{TG} + \mathcal{B}_N+ \mathcal{A}_{N}\cos\left(kL\right)
    =\dfrac{N}{N-1}\mathcal{C}_{N}^\mathrm{TG}+ \mathcal{A}_{N}\cos\left(kL\right),
    \label{CTG}
 \end{equation}
where $\mathcal{C}_N^\mathrm{TG}$ is the Tan contact related to this system, and $\mathcal{B}_N$ and $\mathcal{A}_N$ are the same for the case of spinless fermions (see Sec.~\ref{A1}).  The similarities between Eqs.~\eqref{CF} and~\eqref{CTG} can be understood from the fact that $\rho_1^{TG}(x, y)$ can always be written in terms of the fermionic reduced density matrices $\rho_{j}^F(x, ..)$  as~\cite{lenard_momentum_1964}
\begin{equation}
      \rho_1^\mathrm{TG}(x,y)=\rho_1^\mathrm{F}(x,y)+\sum_{j=1}^{N-1}\dfrac{(-2)^j}{j!}\int_x^{y}\!\!dx_{2}\dots dx_{1+j}\,\rho_{1+j}^\mathrm{F}(x,x_2,\dots;y,x_2,\dots),
    \label{lenseriesmain1}
\end{equation}
which is Eq.~(5) of the main text. 

The first term of Eq.~\eqref{lenseriesmain1} gives rise to the same terms of Eq.~\eqref{CF}. We now show why there is not a $(-1)^{N-1}$ factor for TG gas with respect to the spinless fermion case. To do that, we  write the integral of the second term in Eq.~\eqref{lenseriesmain1} as follows:
\begin{equation}
\begin{split}
     \int_{-L/2}^{L/2}\!\! dx_{2}\dots dx_{1+j}\,\rho_{1+j}^\mathrm{F}(x,x_2,\dots;y,x_2,\dots)=(N-1)(N-2)\dots(N-j)\rho_{1}^\mathrm{F}(x,y) = \frac{(N-1)!}{(N-j-1)!}\rho_{1}^\mathrm{F}(x,y),
     \end{split}
     \label{eq:rho1+j_1}
\end{equation}
and we notice that
\begin{equation}
\sum_{j=1}^{N-1}\dfrac{(-2)^j}{j!}\dfrac{(N-1)!}{(N-j-1)!}=\sum_{j=1}^{N-1}{(-2)^j}{\begin{pmatrix}N-1\\j\end{pmatrix}}=\sum_{j=0}^{N-1}{(-2)^j}{\begin{pmatrix}N-1\\j\end{pmatrix}}-1
=(-1)^{N-1}-1.
\label{eq:sum}
\end{equation}
Using Eq.~(\ref{eq:rho1+j_1}) and~\eqref{eq:sum}, we then evaluate 
\begin{equation}
\begin{split}
\lim_{\substack{x\rightarrow -\frac{L}{2}\\ y\rightarrow \frac{L}{2}}}\sum_{j=1}^{N-1}\dfrac{(-2)^j}{j!}\int_x^{y}\!\!dx_{2}\dots dx_{1+j}\,\rho_{1+j}^{F}(x,x_2,\dots;y,x_2,\dots)
         &=\sum_{j=1}^{N-1}\dfrac{(-2)^j}{j!}\dfrac{(N-1)!}{(N-j-1)!}\rho_{1}^\mathrm{F}(x,y)|_{{x\sim -\frac{L}{2}, y\sim frac{L}{2}}},\\
         &=\rho_{1}^\mathrm{F}(x,y)|_{x\sim -\frac{L}{2}, y\sim \frac{L}{2}} [(-1)^{N-1} -1],
         \end{split}
         \label{pat-gian}
\end{equation}
which gives a factor $-2$ for $N$ even and $0$ for $N$ odd (cf. Eq.~(11) of the main text). 
The sum of Eq.~\eqref{pat-gian} and the first term in the expansion in Eq.~\eqref{lenseriesmain1} appears then in Eq.~\eqref{CTG} with a positive sign in front of $\mathcal{A}_N$ regardless of the parity of $N$.

We now evaluate the additional term, which corresponds to the Tan contact.  This term differs from the others because it is connected to two-body correlations. Indeed, it can be written as~\cite{santana_scaling_2019}
\begin{equation}
   \mathcal{C}_{N}^\mathrm{TG}=\dfrac{2}{\pi} \int_{-L/2}^{L/2} dx_2 \lim_{x,y\rightarrow x_2} \dfrac{\rho_{2}^\mathrm{F}(x,x_2;y,x_2)}{|x-x_2||y-x_2|} =\dfrac{(N-1)N(N+1)(2N+1)}{3L^3}\pi.
\end{equation}
Finally, we see that, for the case of a TG gas, $\mathcal{C}_{N}^\mathrm{TG}=(N-1) \mathcal{B}_{N}$, and, therefore, we  end up with Eq.~\eqref{CTG}.

\subsection{Mixtures}
\label{sec:gen_cas}
As a reminder, the large-$k$ tail of $n(k)$ for $N$-particle mixtures with $\kappa$ spin components  (Eq. (18) of the main text) is given by
\begin{equation}
    \lim_{k\rightarrow\infty}k^4n^\mathrm{mix}(k)=\mathcal{C}^\mathrm{mix}_{N}\!+\!\mathcal{B}_{N}\!+\!(-1)^{N\!+\!1}\!\!\mathcal{A}_{N}\sum_{\sigma}\frac{N_\sigma}{N} c_{\sigma}^{(1,N)}\!\!\cos(kL)
    =\dfrac{S\!+\!1}{N\!-\!1}\mathcal{C}_N^\mathrm{TG}\! +\!(-1)^{N\!+\!1}\!\!\mathcal{A}_{N}\sum_{\sigma}^\kappa\frac{N_\sigma}{N} c_{\sigma}^{(1,N)}\!\!\cos\left(kL\right),\label{CMix}
\end{equation}
where $\mathcal{C}_N^\mathrm{mix}$ and $\mathcal{C}_N^\mathrm{TG}$ are the Tan contacts for the mixture and the Tonks-Girardeau gas, respectively, $\mathcal{B}_N$ and $\mathcal{A}_N$ are the same constants of the spinless fermions (see Sec.~\ref{A1}), $c_\sigma^{(1, N)}$ is the spin correlation related to the cyclic permutation $P_{1 \to N}$ (see Eq. (17) of the main text), $N_\sigma$ is the number of particles with spin $\sigma$, and $S$ is defined in  Eq.~(19) of the main text. 

As discussed in the main text, the one-body density matrix for a mixture  can be written as 
\begin{equation}
\rho_{1}(x,y)=\sum_\sigma^\kappa N_\sigma \rho_{1,\sigma}(x,y)\,\,\,\,\,\,\text{with}\,\,\,\,\,\,
   \rho_{1,\sigma}(x,y)=\sum_{i,j=1}^N c^{(i,j)}_{\sigma}\rho^{(i,j)}(x,y),
    \label{rho1_mix}
\end{equation}
where  
  $c^{(i,j)}_{\sigma}$ and $\rho^{(i,j)}(x,y)$ are the spin and spatial parts calculated on the sector $x_{1,\sigma_1}<\dots<x_{i-1,\sigma_{i-1}}<x<x_{i+1,\sigma_{i+1}}<\dots<x_{j,\sigma_j}<y<x_{j+1,\sigma_{j+1}}<\dots<x_{N,\sigma_N}$~\cite{deuretzbacher_momentum_2016}. The definition of $c^{(i,j)}_{\sigma}$ is given in Eq. (17) of the main text.
We now consider $\rho^{(i, j)}(x,y)$,
which is defined for $x<y$ as~\cite{deuretzbacher_momentum_2016}
\begin{align}
\rho^{(i,j)}(x,y)=\theta(x,y)N!\int_{x_1<\dots<x_{i-1}<x<x_{i+1}<\dots<x_j<y<x_{j+1}<\dots<x_N}
~~dx_2\dots dx_{N}\Psi_A^*(x,x_2,\dots,x_N)\Psi_A(y,x_2,\dots,x_N).
\label{eq:rhoij_orb}
\end{align}

For completeness, the $y<x$ part of Eq.~(\ref{eq:rhoij_orb}) can be obtained using the symmetry relation $\rho^{(i,j)}(x,y)=\rho^{(j,i)}(y,x)$. It must be pointed out that the product $\Psi_A^*(x_1,\cdots,x_{i-1},x,x_{i+1},\dots,x_N)\Psi_A(x_1,\cdots,x_{j-1},y,x_{j+1},\dots,x_N)=\Psi_A^*(x,x_2,\dots,x_N)\Psi_A(y,x_2,\dots,x_N)$, $\forall i,j$, because $\Psi_A$ has the form of a determinant.
For hard-core particles in a box of length $L$, we can rewrite Eq.~\eqref{eq:rhoij_orb} as
\begin{align}
\rho^{(i,j)}(x,y)=\dfrac{ \theta(x,y)N!}{(i-1)!(j-i)!(N-j)!}\int_{-L/2}^x
~~dx_2\dots dx_{i}\int_x^y dx_{i+1}\dots dx_{j}\int_y^{L/2} dx_{j+1}\dots dx_{N}\Psi_A^*(x,x_2,\dots,x_N)\Psi_A(y,x_2,\dots,x_N),
\label{eq:rhoij}
\end{align}

where the wave function $\Psi_A$ is the fully anti-symmetric fermionic wave function:
\begin{equation}
    \Psi_A=\dfrac{1}{\sqrt{N!}}\det\left[\phi_m(x_n)\right],
    \label{eq:PsiA}
\end{equation}
where $\phi_m(x_n)=\sqrt{2/L}\,\sin{[k_m (x_n+L/2)]}$ are natural orbitals  with $k_m=m\pi/L$, $x_n\in[-L/2,L/2]$ and $n,m\in\{1,\dots,N\}$.

Following Refs.~\cite{deuretzbacher_momentum_2016, decamp_exact_2016, decamp_strong_2017}, we use the Leibniz formula for a determinant such that $\Psi_A =(1/\sqrt{N!})\sum_{P\in S_N} \epsilon(P) \prod_{i=1}^{N} \phi_{P(i)}(x_i)$, where $\epsilon(P)$ is the signature of the permutation $P$, to express Eq.~\eqref{eq:rhoij} as
\begin{equation}
    \rho^{(i,j)}(x,y)=\dfrac{1}{(i-1)!(j-i)!(N-j)!}\theta(x,y)\sum_{P,Q\in S_N}\epsilon(P)\epsilon(Q)\phi_{P(1)}(x)\phi_{Q(1)}(y)\prod_{k=2}^{N}\int_{L_{ij}(k)}^{U_{ij}(k)}\phi_{P(k)}(z)\phi_{Q(k)}(z),
    \label{rhoij leibniz}
\end{equation}   
where the integration intervals are defined as
\begin{equation}
    (L_{ij}(k),U_{ij}(k))=\begin{cases}(y,L/2)&\text{if$~j\leq k$,}\\
(x,y)&\text{if$~i\leq k< j$,}\\
(-L/2,x)&\text{if$~k< i$.}\end{cases}\label{CdtLeibniz}
\end{equation}
We then define, for convenience,
\begin{align}
X_{p,q}(x,y)=\phi_{p}(x)\phi_{q}(y),&\\
A_{p,q}(z)=\int_{z}^{L/2}du~\phi_{p}(u)\phi_{q}(u)&
=\dfrac{\sin\left(\dfrac{\pi}{2L}(p+q)(2z+L)\right)}{\pi(p+q)}-\dfrac{\sin\left(\dfrac{\pi}{2L}(p-q)(2z+L)\right)}{\pi(p-q)}.
\end{align}
and re-write Eq.~\eqref{rhoij leibniz} as follows:
\begin{equation}
\begin{split}
    \rho^{(i,j)}(x,y)&=\dfrac{1}{(i-1)!(j-i)!(N-j)!}\theta(x,y)\sum_{P,Q\in S_N}\epsilon(P)\epsilon(Q)X_{P(1),Q(1)}(x,y)\\&\times\prod_{k=2}^{i}( \delta_{P(k), Q(k)}-A_{P(k),Q(k)}(x))\prod_{l=i+1}^{j}\left(A_{P(l),Q(l)}(x)-A_{P(l),Q(l)}(y)\right)\prod_{m=j+1}^{N}A_{P(m),Q(m)}(y),
\end{split}
 \label{rhoij MmP}
\end{equation}
where we have used that the integrals in the three intervals defined in Eq.~\eqref{CdtLeibniz} can be written all in terms of $A_{p,q}(z)$.  Eq.~\eqref{rhoij MmP} represents the starting point of the derivation of every term in Eq.~\eqref{CMix}, which we will consider separately in the following paragraphs. 

\paragraph*{The $k$-independent term.}
Following Sec.~\ref{A1}, we first evaluate Eq.~\eqref{rhoij MmP} in the limit $(x,y)\rightarrow(\pm L/2,\pm L/2)$. To do so, we first notice that
\begin{align}
    &\lim_{\substack{x\rightarrow -\frac{L}{2}\\ y\rightarrow -\frac{L}{2}}}X_{p,q}(x,y)=\dfrac{2\pi^2}{L^3}pq\left(x+\dfrac{L}{2}\right)\left(y+\dfrac{L}{2}\right),\label{Xlim cte}\\
    &\lim_{\substack{z\rightarrow -\frac{L}{2}}} A_{p,q}(z)=\delta_{p,q},\quad \text{and} \quad \lim_{\substack{z\rightarrow \frac{L}{2}}} A_{p,q}(z)= 0 \,\,\forall p, q.\label{Alim}
\end{align}
Therefore, the only nonzero terms in Eq.~\eqref{rhoij MmP} correspond to $i=j=1$ for the case $(x, y) \to -L/2$ and $i=j=N$  for the case $(x, y) \to L/2$. Using Eq.~(\ref{rho1_mix}), 
we can write 
\begin{equation}
   \mathcal{B}_N = \dfrac{1}{2\pi}\left(\lim_{\substack{x\rightarrow -\frac{L}{2}\\ y\rightarrow -\frac{L}{2}}}
    \dfrac{\rho^{(1,1)}(x,y)}{|x+\frac{L}{2}||y+\frac{L}{2}|}\sum_\sigma N_\sigma c^{(1,1)}_\sigma+\lim_{\substack{x\rightarrow +\frac{L}{2}\\ y\rightarrow +\frac{L}{2}}}
    \dfrac{\rho^{(N,N)}(x,y)}{|x-\frac{L}{2}||y-\frac{L}{2}|}\sum_\sigma N_\sigma c^{(N,N)}_\sigma\right)
    =\dfrac{N}{\pi}\lim_{\substack{x\rightarrow -\frac{L}{2}\\ y\rightarrow -\frac{L}{2}}}
    \dfrac{\rho^{(1,1)}(x,y)}{|x+\frac{L}{2}||y+\frac{L}{2}|},
    \label{CFMix}
\end{equation}
 where we have used that $\rho^{(1, 1)}(x, y)|_{(x, y) \to (- \frac{L}{2},-\frac{L}{2})} = \rho^{(N, N)}(x, y)|_{(x, y) \to (\frac{L}{2},\frac{L}{2})}$ and $N=\sum_\sigma N_\sigma c_\sigma^{(i, i)}= \sum_\sigma N_\sigma$ $\forall i$, and
\begin{equation}
    \rho^{(1,1)}(x,y)=\dfrac{1}{(N-1)!}\theta(x,y)\sum_{P,Q\in S_N}\epsilon(P)\epsilon(Q)X_{P(1),Q(1)}(x,y)\prod_{k=2}^{N}A_{P(k),Q(k)}(y).
    \label{eq:rho11}
\end{equation}
Using Eqs.~\eqref{Xlim cte} and~\eqref{Alim}, we can replace $A_{P(k),Q(k)}(y)$ with $\delta_{P(k),Q(k)}$
and obtain
\begin{equation}
    \lim_{\substack{x\rightarrow -\frac{L}{2}\\ y\rightarrow -\frac{L}{2}}}
    \dfrac{\rho^{(1,1)}(x,y)}{|x+\frac{L}{2}||y+\frac{L}{2}|}=
    \dfrac{1}{(N-1)!}\dfrac{2\pi^2}{L^3}\sum_{P\in S_N}P(1)^2
    =\dfrac{(N-1)!}{N(N-1)!}\dfrac{2\pi^2}{L^3}\sum_{n=1}^Nn^2=\dfrac{(N+1)(2N+1)}{3 L^3}\pi^2, \label{eq:limrho11}
\end{equation}
 where we have used the properties of the permutation group.

Finally, by inserting Eq.~\eqref{eq:limrho11} in  Eq.~\eqref{CFMix}, we find
\begin{align}
    \mathcal{B}_N = \dfrac{N(N+1)(2N+1)}{3L^3}\pi,
\end{align}
 which is exactly Eq.~\eqref{pat1}. Therefore, we have shown how to derive Eq.~\eqref{pat1} by starting from this more general problem.

\paragraph*{The Tan contact term.}
The Tan contact term does not depend on the finite size of the system and can be derived by standard methods (see, for example, Refs.~\cite{decamp_high-momentum_2016, Aupetit2022}). For completeness, here we give a few details of the derivation. One can start from the Tan relation 
\begin{equation}
    \mathcal{C}^\mathrm{mix}_{N} = -\dfrac{m^2}{\pi\hbar^4}\frac{\partial E}{\partial (1/g)} \Big|_{g\rightarrow\infty} = \dfrac{S \alpha_N}{\pi},
    \label{eq:CNmix}
\end{equation}
where the total energy for the mixture in the limit $g\to \infty$ can be written as $E \simeq E_A - m^2 S\alpha_N/(\hbar^4g)$ with
\begin{equation}
    \alpha_N=\frac{N!}{(N-2)!}\int_{-L/2}^{L/2}dx_1\left(\prod_{i=1,i\neq2}^{N-1}\int_{x_{i}}^{L/2}dx_{i+1}\right)\left|\frac{\partial\Psi_A}{\partial x_1}\right|^2= \frac{N(N+1)(2N+1) \pi^2}{6 L^3},
    \label{alpha box}
\end{equation}
and 
\begin{equation}
  S=\sum_{P}\sum_{i=1}^{N-1}\left[\frac{1}{4}(a_P-a_{P_{i,i+1}})^2(1- \delta^{\sigma_{i+1}}_{\sigma_i})+\eta a_{P}a_{P_{i, i+1}}\delta^{\sigma_{i+1}}_{\sigma_i}\right],
\end{equation}
with $\eta$ is equal to 1 for bosons and 0 for fermions. One then notices that $\alpha_N = \mathcal{B}_N \pi/2$ and, using that $\mathcal{C}_N^\mathrm{TG} = (N-1) \mathcal{B}_N$ (Eq. (10) of the main text), one can collect the two constant terms as
\begin{equation}
\mathcal{C}^\mathrm{mix}_{N} + \mathcal{B}_N = \frac{S}{N-1}\mathcal{C}_N^\mathrm{TG} + \frac{\mathcal{C}_N^\mathrm{TG}}{(N-1)} = \frac{S+1}{N-1} \mathcal{C}_N^\mathrm{TG}.
    \label{eq:CNTGBN}
\end{equation}

\paragraph*{The oscillating term.}
We now evaluate the amplitude of the oscillation in Eq.~\eqref{CMix}. As discussed in Sec.~\ref{A1}, this derives from the contribution at the borders of the trap, namely, by taking the limit $(x,y)\rightarrow(\mp L/2, \pm L/2)$. To do so, we first notice that 
\begin{align}
    &\lim_{\substack{x\rightarrow -\frac{L}{2}\\ y\rightarrow \frac{L}{2}}}X_{p,q}(x,y)=\dfrac{2\pi^2}{L^3}(-1)^qpq\left(x+\dfrac{1}{2}\right)\left(y-\dfrac{1}{2}\right),\label{Xlim osc}\\
    &\lim_{\substack{x\rightarrow -\frac{L}{2}\\ y\rightarrow \frac{L}{2}}}\left(A_{p,q}(x)-A_{p,q}(y)\right)=\delta_{p,q}.\label{AmAlim}
\end{align}
Thanks to these limits and using the symmetries of $\rho^{(i,j)}(x, y)$, we see that the only nonzero terms in Eq.~\eqref{rhoij MmP} correspond to $(i, j)=(1, N)$ for $(x, y) \to (-L/2, L/2)$ and $(i,j)=(N, 1)$ for $(x, y) \to (L/2, -L/2)$.
Following Sec.~\ref{A1}, we can define \begin{equation}
\begin{split}
(-1)^{N+1} \mathcal{A}_N^\mathrm{mix} &=
    \dfrac{1}{2\pi}\left(\lim_{\substack{x\rightarrow -\frac{L}{2}\\ y\rightarrow +\frac{L}{2}}}
    \dfrac{\rho^{(1,N)}(x,y)}{|x+\frac{L}{2}||y-\frac{L}{2}|}\sum_\sigma N_\sigma c^{(1,N)}_\sigma+\lim_{\substack{x\rightarrow +\frac{L}{2}\\ y\rightarrow -\frac{L}{2}}}
    \dfrac{\rho^{(N,1)}(x,y)}{|x-\frac{L}{2}||y+\frac{L}{2}|}\sum_\sigma N_\sigma c^{(N,1)}_\sigma\right)\\&=\dfrac{1}{\pi}\left(\lim_{\substack{x\rightarrow -\frac{L}{2}\\ y\rightarrow +\frac{L}{2}}}
    \dfrac{\rho^{(1,N)}(x,y)}{|x+\frac{L}{2}||y-\frac{L}{2}|}\right)\sum_\sigma N_\sigma c^{(1,N)}_\sigma,
    \end{split}
     \label{CmixOsc}
\end{equation}
 where we have used the symmetry properties $c_\sigma^{(1, N)} = c_\sigma^{(N, 1)}$ and $\rho^{(1, N)}(x, y) = \rho^{(N, 1)}(y, x)$ and that
\begin{align}
    \rho^{(1,N)}(x,y)=\dfrac{1}{(N-1)!}\sum_{P,Q\in S_N}\epsilon(P)\epsilon(Q)X_{P(1),Q(1)}(x,y)\prod_{l=2}^{N}\left(A_{P(l),Q(l)}(x)-A_{P(l),Q(l)}(y)\right).
\end{align}
 Using Eqs.~\eqref{Xlim osc}, \eqref{AmAlim} and~\eqref{Alim}, we can therefore replace the difference $A_{p, q}(x)- A_{p, q}(y)$ with a $\delta_{p, q}$, which corresponds to have $P=Q$, and obtain

\begin{align}
    &\lim_{\substack{x\rightarrow -\frac{L}{2}\\ y\rightarrow \frac{L}{2}}}
   \dfrac{\rho^{(1,N)}(x,y)}{|x+\frac{L}{2}||y-\frac{L}{2}|}
   =\dfrac{1}{(N-1)!}\dfrac{2\pi^2}{L^3}\sum_{P\in S_N}(-1)^{P(1)}P(1)^2
    =\dfrac{2\pi^2}{L^3}\dfrac{(N-1)!}{N(N-1)!}\sum_{n=1}^N(-1)^{n}n^2
    =\dfrac{\pi^2}{L^3}(-1)^{N+1}(N+1),&\label{N-1 osc}
\end{align}
where we have used properties of the group of permutations, as done in Eq.~\eqref{eq:limrho11}.
Finally, we find a more general definition of the amplitude of the oscillation, namely,
\begin{equation}
    (-1)^{N+1}\mathcal{A}_N^\mathrm{mix}= (-1)^{N+1}\mathcal{A}_N  \sum_\sigma \dfrac{N_\sigma}{N}  c^{(1,N)}_\sigma,
    \label{eq:Amix}
\end{equation}
which is consistent with Eq.~\eqref{pat2} and takes into account the presence (absence) of the factor $(-1)^{N+1}$ for fermions (bosons). Indeed, as explained in the main text, $c_\sigma^{(1, N)}=1$ for fermions and   $(-1)^{N+1}$ for Tonks-Girardeau bosons, whose wave function can be written under the form of symmetrized Slater determinant.

\section{Example of $c^{(1,N)}_\sigma$ calculation}
\label{sec:c1N}
In this section, we detail the construction of the quantity $c_\sigma^{(1,N)}$ for the case of a balanced mixture of $N=4$ bosons with spin components $\sigma=\uparrow$ or $\downarrow$. 
The correlation $c_\sigma^{(1, N)}$ corresponds to the $(1, N)$ element of the spin part of the one-body density matrix  $c_\sigma^{(i, j)}$ (see Eq.~\eqref{rho1_mix}), which is
\begin{equation}
    c_\sigma^{(i,j)}=\delta_{\sigma}^{\sigma_i}\sum_{P\in S_N} a_{P} a_{P_{i\rightarrow j}}, 
    \label{eq:cij}
\end{equation}
where $\delta_{\sigma}^{\sigma_i}$ selects only the sites with spin $\sigma_i=\sigma$ and $a_{P_{i\rightarrow j}}$ is the sector coefficient obtained by starting from the spin configuration related to $a_P$ and applying a cyclic permutation, which takes the $i$-th into the $j$-th position, and vice versa. As derived in Sec.~\ref{sec:gen_cas}, $c_\sigma^{(1,N)}$ contributes to the large-$k$ behavior of $n(k)$ for a mixture.

Let us start considering the $\uparrow$ spin-component. The $a_P$ coefficients that are involved in the calculation of $c^{(1,4)}_\uparrow$ have been gathered in Table \ref{table:ap2p2up}.
\begin{table}[ht]
\caption{Coefficients $a_i$ corresponding to the different spin sectors $P$ and $P_{1\rightarrow 4}$ with the selection rule $\delta_{\uparrow,\sigma_1}$, for the case of a $2+2$ bosonic mixture.}
\centering 
\begin{tabular}{c c c c} 
\hline \hline
\rule{0pt}{2.5ex} Sector $P$ $\delta_{\uparrow, \sigma_1}$ &$\quad\quad a_P\quad\quad$ & Sector $P_{1\rightarrow 4}$ $\delta_{\uparrow, \sigma_1}$& $\quad\quad a_{P_{1\rightarrow 4}}\quad\quad$   \\ [0.5ex]
\hline 
$x_{1,\uparrow}<x_{2,\uparrow}<x_{3,\downarrow}<x_{4,\downarrow}$  & $a_1$  & $x_{2,\uparrow}<x_{3,\downarrow}<x_{4,\downarrow}<x_{1,\uparrow}$  &$-a_3$	\\
$x_{1,\uparrow}<x_{2,\uparrow}<x_{4,\downarrow}<x_{3,\downarrow}$  & $-a_1$ & $x_{2,\uparrow}<x_{4,\downarrow}<x_{3,\downarrow}<x_{1,\uparrow}$  &$a_3$ \\ 
$x_{1,\uparrow}<x_{3,\downarrow}<x_{2,\uparrow}<x_{4,\downarrow}$  & $a_2$  & $x_{3,\downarrow}<x_{2,\uparrow}<x_{4,\downarrow}<x_{1,\uparrow}$  &$-a_5$	\\
$x_{1,\uparrow}<x_{4,\downarrow}<x_{2,\uparrow}<x_{3,\downarrow}$  & $-a_2$  & $x_{4,\downarrow}<x_{2,\uparrow}<x_{3,\downarrow}<x_{1,\uparrow}$  &$a_5$ \\
$x_{1,\uparrow}<x_{3,\downarrow}<x_{4,\downarrow}<x_{2,\uparrow}$  & $a_3$ & $x_{3,\downarrow}<x_{4,\downarrow}<x_{2,\uparrow}<x_{1,\uparrow}$  &$-a_6$ \\ 
$x_{1,\uparrow}<x_{4,\downarrow}<x_{3,\downarrow}<x_{2,\uparrow}$ & $-a_3$ & $x_{4,\downarrow}<x_{3,\downarrow}<x_{2,\uparrow}<x_{1,\uparrow}$ & $a_6$ \\ [1ex] 
\hline
\end{tabular}
\label{table:ap2p2up}
\end{table}
The coefficients $\{a_1,\dots,a_6\}$ in Table \ref{table:ap2p2up} are related to the sectors collected in the snippet basis $\{\uparrow\uparrow\downarrow\downarrow,\uparrow\downarrow\uparrow\downarrow,\uparrow\downarrow\downarrow\uparrow,\downarrow\uparrow\uparrow\downarrow,\downarrow\uparrow\downarrow\uparrow,\downarrow\downarrow\uparrow\uparrow\}$.
We remind that the snippet basis has $N!/(N_\uparrow ! N_\downarrow !)$ elements instead of $N!$, which are all the possible permutations of $N$ particles. However, because we use as a basis the antisymmetric fermionic many-body wave function $\Psi_A$, we have to include a minus sign if we switch two bosons (see, for instance, the first two rows of Table~\ref{table:ap2p2up}).

Using Eq.~(\ref{eq:cij}), one obtains
\begin{equation}               
    c^{(1,4)}_\uparrow=-2a_1a_3-2a_2a_5-2a_3a_6.
    \label{c-up}
\end{equation}

\begin{table}[ht]
\caption{Coefficients $a_i$ corresponding to the different spin sectors $P$ and $P_{1\rightarrow 4}$ with the selection rule $\delta_{\downarrow,\sigma_1}$, for the case of a $2+2$ bosonic mixture.}
\centering 
\begin{tabular}{c c c c} 
\hline \hline
\rule{0pt}{2.5ex} Sector $P$ $\delta_{\downarrow, \sigma_1}$ &$\quad\quad a_P\quad\quad$ & Sector $P_{1\rightarrow 4}$ $\delta_{\downarrow, \sigma_1}$& $\quad\quad a_{P_{1\rightarrow 4}}\quad\quad$   \\ [0.5ex]
\hline 
$x_{3,\downarrow}<x_{1,\uparrow}<x_{2,\uparrow}<x_{4,\downarrow}$  & $a_4$  & $x_{1,\uparrow}<x_{2,\uparrow}<x_{4,\downarrow}<x_{3,\downarrow}$  &$-a_1$	\\
$x_{3,\downarrow}<x_{2,\uparrow}<x_{1,\uparrow}<x_{4,\downarrow}$  & $-a_4$ & $x_{2,\uparrow}<x_{1,\uparrow}<x_{4,\downarrow}<x_{3,\downarrow}$  &$a_1$ \\ 
$x_{3,\downarrow}<x_{1,\uparrow}<x_{4,\downarrow}<x_{2,\uparrow}$  & $a_5$  & $x_{1,\uparrow}<x_{4,\downarrow}<x_{2,\uparrow}<x_{3,\downarrow}$  &$-a_2$	\\
$x_{3,\downarrow}<x_{2,\uparrow}<x_{4,\downarrow}<x_{1,\uparrow}$  & $-a_5$  & $x_{2,\uparrow}<x_{4,\downarrow}<x_{1,\uparrow}<x_{3,\downarrow}$  &$a_2$ \\
$x_{3,\downarrow}<x_{4,\downarrow}<x_{1,\uparrow}<x_{2,\uparrow}$  & $a_6$ & $x_{4,\downarrow}<x_{1,\uparrow}<x_{2,\uparrow}<x_{3,\downarrow}$  &$-a_4$ \\ 
$x_{3,\downarrow}<x_{4,\downarrow}<x_{2,\uparrow}<x_{1,\uparrow}$ & $-a_6$ & $x_{4,\downarrow}<x_{2,\uparrow}<x_{1,\uparrow}<x_{3,\downarrow}$ & $a_6$ \\ [1ex] 
\hline
\end{tabular}
\label{table:ap2p2down}
\end{table}

Analogously, using Table \ref{table:ap2p2down}, one can show that
\begin{equation}               
    c^{1,4}_\downarrow=-2a_4a_1-2a_5a_2-2a_6a_2.
    \label{c-down}
\end{equation}
For convenience, we define the phase of the oscillations of $\mathcal{K}_N^{\text{mix}}$ (see Eq.~(18) of the main text) as
\begin{equation}
\Phi_N^\mathrm{mix}=(-1)^{N+1}\sum_\sigma \dfrac{N_\sigma}{N}  c^{(1,N)}_\sigma=(-1)^{N+1}\frac{\mathcal{A}_N^\mathrm{mix}}{\mathcal{A}_N},
\label{eq:Phimix}
\end{equation}
where $\mathcal{A}_N^\mathrm{mix}$ is defined in Eq.~\eqref{eq:Amix}. 
For the mixture of $2+2$ bosons, Eq.~\eqref{eq:Phimix} becomes
\begin{equation}
\Phi_4^{2+2}=a_1a_3+a_1a_4+2a_2a_5+a_2a_6+a_3a_6.
    \label{eq:Phi4_2+2}
\end{equation}
Due to the specific permutation rule present in $c_\sigma^{(1, N)}$, Eq.~\eqref{eq:Phi4_2+2} clearly involves only certain products of $a_i a_j$ with $i \neq j$. This means that, for example, there is no possibility to connect the snippet $\uparrow\downarrow\downarrow\uparrow$ corresponding to the coefficients $\pm a_3$ to the snippet $\downarrow\uparrow\uparrow\downarrow$ corresponding to the coefficients $\pm a_4$.
This justifies the fact that the third excited state of the $SU(2)$ $2+2$ bosonic mixture (see Fig.~2 in the main text) shows no oscillations in the tails of the momentum distribution. Indeed, the spin part of the many-body wave function for this state is given by $|\uparrow\downarrow\downarrow\uparrow \rangle -  |\downarrow\uparrow\uparrow\downarrow \rangle$ (up to a normalization factor) and, consequently, $\Phi_4^{2+2}$ is zero. 

\subsection{Consideration on the behavior of $\Phi_N^\mathrm{mix}$}
The phase $\Phi_N^\mathrm{mix}$ depends on the number of particles $N$, on the type of mixture, and on the system state (ground-state or excited states).
Indeed, it can be positive or negative, with a magnitude included between $0$ and $1$.
This is shown in Table~\ref{table:apsN4M2}, where we have gathered the values
of $\Phi_N^\mathrm{mix}$ for the case of $SU(2)$ bosonic and fermionic mixtures for a different number of total particles and particles per component.

\begin{table}[H]
\caption{Phase of the oscillations $\Phi_N^\mathrm{mix}$ (Eq.~\eqref{eq:Phimix}) as a function of spin excitation state  for the case of $SU(2)$ bosonic mixtures (central columns) and fermionic mixtures (right columns) with $1+4$, $2+2$, $2+3$, and $3+3$ particles. ``GS'' stands for Ground State and ``MS'' for Most excited State of the lowest energy manifold.}
\label{table:apsN4M2}
\centering 
\begin{tabular}{c | c c c c | c c c c c} 
\hline \hline
\rule{0pt}{2.5ex} Excitation level & $1B\!+\!4B$ & $2B\!+\!2B$ & $2B\!+\!3B$ & $3B\!+\!3B$ & $1F\!+\!4F$ & $2F\!+\!2F$ & $2F\!+\!3F$ & $3F\!+\!3F$ & \\ [0.5ex]
\hline 
GS & $1$ & $1$ & $1$ & $1$& $-0.81$& $-0.87$& $0.33$& $0.78$ &\\
$1$ & $0.09$ & $-0.15$ & $0.08$ & $0.24$& $-0.6$& $0.85$& $0.28$& $-0.74$ &\\
$2$ & $0.31$ & $-0.87$ & $-0.6$ & $-0.37$& $0.31$& $0$& $-0.81$& $0.18$ &\\
$3$ & $-0.59$ & $0$ & $0.31$ & $-0.81$& $0.09$& $0.87$& $0.76$& $-0.77$ &\\
$4$ & $-$ & $-0.85$ & $-0.78$ & $0.5$& $-$& $0.15$& $-0.58$& $-0.2$ &\\
$5$ & $-$ & $-$ & $-0.58$ & $-0.47$& $-$& $-$& $-0.78$& $-0.07$ &\\
$6$ & $-$ & $-$ & $0.76$ & $-0.78$& $-$& $-$& $0.31$& $-0.5$ &\\
$7$ & $-$ & $-$ & $-0.81$ & $-0.33$& $-$& $-$& $-0.6$& $0.91$ &\\
$8$ & $-$ & $-$ & $0.28$ & $0.63$& $-$& $-$& $0.09$& $-0.15$ &\\
$9$ & $-$ & $-$ & $-$ & $0.32$& $-$& $-$& $-$& $0.5$ &\\
$10$ & $-$ & $-$ & $-$ & $-0.5$& $-$& $-$& $-$& $-0.32$ &\\
$11$ & $-$ & $-$ & $-$ & $0.15$& $-$& $-$& $-$& $-0.63$ &\\
$12$ & $-$ & $-$ & $-$ & $-0.91$& $-$& $-$& $-$& $0.33$ &\\
$13$ & $-$ & $-$ & $-$ & $0.5$& $-$& $-$& $-$& $0.78$ &\\
$14$ & $-$ & $-$ & $-$ & $0.07$& $-$& $-$& $-$& $0.47$ &\\
$15$ & $-$ & $-$ & $-$ & $0.2$& $-$& $-$& $-$& $-0.5$ &\\
$16$ & $-$ & $-$ & $-$ & $0.77$& $-$& $-$& $-$& $0.81$ &\\
$17$ & $-$ & $-$ & $-$ & $-0.18$& $-$& $-$& $-$& $0.37$ &\\
$18$ & $-$ & $-$ & $-$ & $0.74$& $-$& $-$& $-$& $-0.24$ &\\
MS & $-0.81$ & $0.87$ & $0.33$ & $-0.78$& $1$& $-1$& $1$& $-1$ &\\[1ex] 
\hline
\end{tabular}
\end{table}
Thus, in typical experimental set-ups, where the momentum distribution is obtained by averaging over systems with a fluctuating number of particles,
the measured amplitude oscillations $\langle\Phi_N^\mathrm{mix}\mathcal{A}_N\rangle$
will generally vanish, except if the system is a SU(2) bosonic mixture and is prepared mostly in the ground-state.
Indeed, we can observe Table~\ref{table:apsN4M2} that for the GS of $SU(2)$ bosonic mixtures, $\Phi_N^{SU(2)}$ is equal to 1 $\forall N$.
This conclusion can be generalized to the case of $SU(\kappa)$ bosons, $\forall \kappa$, but does not hold if the $SU(\kappa)$ symmetry is broken or for other types of mixtures.